\documentclass[12pt,psfig,epsfig,graphicx,psfrag]{article}
\usepackage{graphicx}
\usepackage{amsfonts,amssymb,amsmath}
\usepackage{hyperref}

\newcommand{\be}{\begin{equation}}
\newcommand{\ee}{\end{equation}}
\newcommand{\ba}{\begin{eqnarray}}
\newcommand{\ea}{\end{eqnarray}}
\newcommand{\tr}{{\mathrm{tr}}}

\def\L5{\tilde{\Lambda}}

\newcommand{\m}{\hat m}
\def\pd{\partial}
\def\a{\alpha}
\def\b{\beta}
\def\kf{\tilde \kappa_f}
\def\g{\gamma}
\def\di{\mathrm{d}}

\def\m{\mu}
\def\n{\nu}

\def\t{\tau}

\def\l{\lambda}

\def\r{\rho}

\def\H{\mathcal H}
\def\L{\Lambda}

\renewcommand{\d}{{\mathrm{d}}}

\def\1{\mathchoice{\rm 1\mskip-4.2mu l}{\rm 1\mskip-4.2mu l}%
{\rm 1\mskip-4.6mu l}{\rm 1\mskip-5.2mu l}}

\begin{document}
\vskip 2cm

\begin{center}
{\Large {\bf Bigravity and Lorentz-violating Massive Gravity}}\\[1cm]
D.~Blas$^{a,}$\footnote{dblas@ffn.ub.es},
C.~Deffayet$^{b,c,}$\footnote{deffayet@iap.fr},
J.~Garriga$^{a,}$\footnote{garriga@ffn.ub.es},
\\
$^a${\it ICC, Departament de F\'isica Fonamental, Universitat de Barcelona,\\
Diagonal 647, 08028 Barcelona, Spain.}\\
$^b${\it APC \footnote{UMR 7164 (CNRS, Universit\'e Paris 7,
CEA, Observatoire de Paris)}, B\^atiment Condorcet, 10 rue Alice Domont et L\'eonie Duquet,\\
 75205 Paris Cedex 13, France.}\\
$^c${\it
GReCO/IAP \footnote{UMR 7095 (CNRS, Universit\'e
Paris 6)}, 98 bis Boulevard Arago, 75014 Paris, France.}
\\
\end{center}
\vskip 0.2cm

\noindent
\begin{abstract}

Bigravity is a natural arena where a non-linear theory of massive gravity can be formulated.
If the interaction between the metrics $f$ and $g$ is non-derivative, spherically symmetric exact solutions can
be found. At large distances from the origin, these are generically Lorentz-breaking bi-flat solutions
(provided that the
corresponding vacuum energies are adjusted appropriately). The spectrum of linearized perturbations around such
backgrounds contains a massless as
well as a massive graviton, with {\em two} physical polarizations each. There are no propagating vectors or scalars, and the theory is ghost free
(as happens with certain massive gravities with explicit breaking of Lorentz invariance). At the linearized level,
corrections to GR are proportional to the square of the graviton mass, and so there is no vDVZ discontinuity.
Surprisingly, the solution of linear theory for a static spherically symmetric source does {\em not} agree with
the linearization of any of the known exact solutions. The latter coincide with the
standard Schwarzschild-(A)dS solutions of General Relativity, with no corrections at all.
Another interesting class of solutions is obtained where $f$ and $g$ are proportional to each other.
The case of bi-de Sitter solutions is analyzed in some detail.

\end{abstract}

\pagebreak

\tableofcontents

\newpage

\section{Introduction}

In  recent years, many proposals to modify gravity at very
large, i.e. cosmological, distances have been put forward with the
hope to find an alternative explanation to the observed acceleration
of the Universe. Indeed, one can imagine that this acceleration is
not due to some, yet unknown and dark, component of the Universe,
but to some failure of General Relativity at large distances (see
e.g. \cite{Albrecht:2006um}).

 To obtain such a modification of gravity, one of the simplest ideas - in fact already present in the original motivation of Einstein to introduce a cosmological constant - is obviously to give some sort of mass to the graviton. However, it has long been known that theories of massive
gravity are rather problematic. In the linearized regime, a Lorentz invariant mass term for the graviton
must have the Fierz-Pauli (FP) form \cite{Fierz:1939ix}, or else ghosts appear in the
spectrum. With the FP mass
term, light deflection caused by static sources differs from that of General Relativity (GR)
even in the limit of massless gravitons, a pathology known as the van Dam-Veltam-Zakharov (vDVZ) discontinuity \cite{vDVZ}.

Among the recently studied proposals to modify gravity at large distances, Dvali-Gabadadze-Porrati gravity (\cite{Dvali:2000hr} DGP gravity)
has been the subject of  numerous works in particular because of its ability to produce a late time acceleration of the expansion of the Universe even with a vanishing cosmological constant \cite{Deffayet:2000uy}. Another theory, sharing with DGP gravity the same tensorial structure of the graviton propagator, and (for that reason) also the vDVZ discontinuity, is non linear massive gravity. The latter theory can be obtained from some bigravity theory, where one of the two metrics is frozen (see \cite{Damour:2002ws} and
references therein). On the other hand, bigravity theories have the ability to produce an accelerated expansion of the Universe in a non standard way \cite{Damour:2002wu}, one of the two metrics being then regarded as some new type of dark energy component
( it
is interesting to note that there is an integration constant appearing in the effective cosmological constant in
these models, which reminds one of unimodular gravity \cite{Weinberg:1988cp}). The vDVZ discontinuity, without a cure, would suffice to rule out any theory in which it appears. It was however proposed that this discontinuity could disappear non perturbatively, i.e. that it would not be seen in exact solutions of the theory, but only shows up as an artifact of the linearization procedure \cite{Vainshtein:1972sx}. While this seems problematic in non linear massive gravity \cite{Jun:1986hg,Damour:2002gp}, there is some evidence that it does work in DGP gravity \cite{Deffayet:2001uk,Gruzinov:2001hp,Dvali:2006if}. It is then of some importance to better understand exact solutions of various types of massive gravities, among which bigravity theories. In this paper we concentrate on spherically solutions of
the latter theories. Indeed, in bigravity theories, non-derivative interactions between
 two different metrics can give a mass to some of the spin-2 polarizations
  and a large class of non trivial exact solutions are known. This is not
  the case for more complicated models such as DGP gravity, and thus bigravity
  theories provides an invaluable framework to investigate various properties of ``massive" gravity and
 related models of large distance modification of gravity.

 Interestingly there is another motivation to take a closer look at bigravity theories. Indeed, it has been shown \cite{Rubakov:2004eb,Dubovsky:2004sg} that the vDVZ
discontinuity can also be avoided when the mass term breaks Lorentz invariance. The different phases of massive gravity
with explicit breaking of Lorentz invariance have been further analyzed in \cite{Dubovsky:2004sg}.
 Some examples with
spontaneous breaking (due to additional vector field condensates) have also been considered
in Ref. \cite{Libanov:2005nv} (see also \cite{Gripaios:2004ms}).
The construction in \cite{Libanov:2005nv} exhibits ghosts and tachyons at low momenta, but it is nevertheless phenomenologically acceptable
for certain choices of parameters. The question of Lorentz violation consistent with
cosmological observations has been recently studied in \cite{Dvali:2007ks}. Such Lorentz violating
mass terms also appear in the so-called ghost condensate scenario with non trivial cosmological
 consequences \cite{ArkaniHamed:2003uy}. It seems worthwhile, in this context, to explore alternative
 scenarios with spontaneous breaking of Lorentz invariance. A straightforward example is provided
  again by spherically symmetric solutions of bigravity theories.

In Ref. \cite{Blas:2005yk} we studied the global properties of
a wide class of bigravity solutions, and here we will develop the theory of linearized perturbations around
some of them. A particularly interesting example is the solution where both metrics $f$ and $g$ are
flat, but with different values of the ``speed of light". This is the simplest case where Lorentz
invariance is spontaneously broken. It is also interesting in its own right, since it corresponds to the far field
in a wider class of spherically symmetric exact solutions, of the Schwarzschild form.

The paper is organized as follows. In Section 2 we review the basics of bigravity.
We show that if the interactions between
the two metrics are non-derivative, then there are always a broad class of solutions in the Schwarzschild-(A)dS
family, with Lorentz-breaking
asymptotics. In Section 3, we study the perturbations to Lorentz breaking bi-flat solutions.
In Section 4 we consider perturbations around solutions where the two metrics are proportional to each other,
concentrating in the case of bi-deSitter solutions. Section 5 summarizes our conclusions.

While this paper was being prepared, a related work \cite{Berezhiani:2007zf} appeared where the study of perturbations to bi-flat Lorentz-breaking
backgrounds is outlined. Where we overlap, our findings agree with those of Ref. \cite{Berezhiani:2007zf}
(see also \cite{Blas:2005P} and \cite{Blas:2005sz}).

\section{Exact Solutions}

Following \cite{Isham:gm}, we consider the action
\begin{equation}
\label{action} S=\int \d^4 x\sqrt{-g}\ \left(\frac{- R_g}{2
\kappa_g} +L_g\right) +\int \d^4 x \sqrt{-f}\ \left(\frac{- R_f}{2
\kappa_f} + L_f\right) + S_{int}[f,g]
\end{equation}
Here $L_f$ and $L_g$ denote generic matter Lagrangians coupled to
the metrics $f$ and $g$ respectively, and subindices $f$ and $g$ on the Ricci scalar $R$ indicate which metric we use to
compute it. For the background solutions, we shall restrict attention to the case where there is
only a vacuum energy term in each matter sector $L_f=-\rho_f,
L_g=-\rho_g$, where $\rho_f$ and $\rho_g$ are constant.
The kinetic terms are invariant under independent diffeomorphisms of the metrics $f$ and $g$,
 but the interaction term is invariant under ``diagonal" diffeomorphisms\footnote{
In principle, we might also include derivative interactions between the two metrics compatible with the diagonal
symmetry, but in general these terms yield a ghost in the vector sector and we
will not consider them here. An interesting possibility would be to preserve
the independent unimodular diffeomorphisms in the kinetic terms, in which case the derivative
coupling is possible.}, under which both metrics transform.

There is much freedom in the choice of the
interaction term in (\ref{action}).
For instance, Ref. \cite{Isham:gm} considered a non-linear generalization of the Fierz-Pauli mass term
\begin{equation}
\label{interaction}
 S_{int}=-\frac{\zeta}{4}\int
\d^4 x
(-g)^u(-f)^v(f^{\mu\nu}-g^{\mu\nu})(f^{\sigma\tau}-g^{\sigma\tau})(g_{\mu\sigma}
g_{\nu\tau}-g_{\mu\nu}g_{\sigma\tau}),
\end{equation}
with \be u+v = \frac{1}{2}.\ee As we shall see in the following
Section, linearized perturbations around asymptotically bi-flat
Lorentz-breaking solutions of this particular theory show a singular
behaviour. For that reason, it will be instructive to look at
generalizations of (\ref{interaction}). The most general interaction
potential which preserves the ``diagonal" diffeomorphism takes the
form
 \cite{Damour:2002ws}
\be
S_{int}=\zeta\int \di^4 x(-g)^u(-f)^{v} V[\{\t_n\}],\label{generi}
\ee
where $\t_n=\tr[{\mathcal M}^n], \ n:1,...,4$ correspond to the traces
of the first four powers of ${\mathcal M}^{\m}_{\n}=f^{\m\a}g_{\a\n}$, and $V$ is an
arbitrary function.

There is also some arbitrariness in the way one introduces matter fields, since one has two different metrics
 at hand. One can  at least couple matter minimally to any one of the two metrics, in which case
 the other metric can be regarded as some kind of exotic new type of matter. Those two choices
 correspond to the two matter Lagrangians $L_g$ and $L_f$, of action (\ref{action}), where it is understood
  that the matter fields entering into $L_g$ and $L_f$ are different. This opens the possibility to have two
   types of matter, one which feels the metric $g$ and the other which feels the metric $f$.
In fact one can imagine more complicated situations in which matter fields would be coupled
to some composite metric built out of the two metrics $f$ and $g$. If one wishes to recover
the standard equivalence principle, one should obviously ask that standard matter only couples
to one metric, and a minimal choice is thus, e.g., that all matter fields appear say in $L_f$
(respectively $L_g$), while $L_g$ (respectively $L_f$), will be simply given by a cosmological constant.
 With such a choice, matter moves along geodesics of the metric $f$ (respectively $g$), and,
  provided the solutions for the metric $f$ are the same as in standard general relativity
  (which turns out to be possible as will be seen below), there would be no deviations from
   4D general relativity seen in matter motion. However, various question arise, should one wish
 to consider bigravity theories as realistic. Among those, the fact that bigravity theories
are likely to suffer from the same type of instability (at the non linear level)
 discovered by Boulware and Deser in non linear massive gravity \cite{Boulware:1973my,Creminelli:2005qk}
 (see however \cite{Damour:2002ws,Damour:2002wu}). Our viewpoint is here more to use those
  theories as a toy to study properties of more complicated models, than pushing the idea
  that bigravity theories are realistic. Thus we will not further discuss in the following
   phenomenological consequences of various types of couplings to matter. In the aim of
   discussing the vDVZ discontinuity, we will only consider
   the possibility to have two types of matter coupled respectively minimally to metrics $f$ and $g$,
 as indicated  in the action (\ref{action}).

Let us then introduce the exact solutions subject of this study.
The general static spherically symmetric ansatz for bi-gravity can be written as \cite{Isham:1977rj}
\ba
\label{formg} g_{\mu \nu} dx^\mu dx^\nu &=&
J dt^2 - K dr^2 - r^2 \left(d \theta^2 + \sin ^2 \theta \; d\phi^2 \right) \\
f_{\mu \nu} dx^\mu dx^\nu &=& C dt^2 - 2 D dt dr - A dr^2 -
B\left(d \theta^2 + \sin ^2\theta \; d \phi^2\right),
\label{formf}
\ea
where the metric coefficients are functions of $r$.
Note that in general it is not possible to write both metrics in diagonal form in the same coordinate system.

For the potential (\ref{interaction}), and
for $D(r)\neq 0$, it was shown in \cite{Isham:1977rj} that the {\em general} solution is given by
\ba
g_{\mu
\nu} \d x^\mu \d x^\nu &=& \left(1-q\right) \d t^2
- (1-q)^{-1} \d r^2 - r^2 (\d \theta^2 + \sin^2 \theta \d \phi^2) \label{spheg}\\
f_{\mu \nu} \d x^\mu \d x^\nu &=& \frac{\g}{\beta} (1-p) \d t^2 -2
D \d t \d r
 - A \d r^2  - \g r^2 (\d\theta^2 + \sin^2 \theta \d\phi^2), \label{sphef}
\ea
where
\ba
A &=&  \frac{\g}{\beta}(1-q)^{-2}\left(p + \beta - q- \beta q \right),\\
\label{Dequation} D^2 &=& \left(\frac{\g}{\beta}\right)^2(1-q)^{-2}(p-q)(p+\beta-1 -\beta q).
\ea
Here
$\beta$ is an arbitrary constant, $\g=2/3$ and the potentials $p$ and $q$
are functions of $r$. Somewhat surprisingly, these potentials turned out to coincide with those of
the standard Schwarzschild-(A)dS family. Solutions of the form (\ref{spheg}-\ref{sphef}) are called Type I.\\

Substituting the Type I ansatz into the effective energy
momentum tensors which are obtained by varying $S_{int}$, one readily finds that
these take the form of cosmological terms:
\ba \label{EQMMOT}
T^f_{\mu \nu} \equiv {-2\over \sqrt{ -f}}
{\delta S_{int}\over \delta f^{\mu\nu}}={\tilde\Lambda_f \over \kappa_f} f_{\mu\nu}, \;\;\;  T^g_{\mu \nu}
\equiv {{-2\over \sqrt{ -g}} {\delta S_{int}\over \delta g^{\mu\nu}}}
={\tilde\Lambda_g\over \kappa_g} g_{\mu\nu},
\ea
where
\ba \frac{\tilde\Lambda_f}{\kappa_f} &=& \frac
{\zeta }{4} \left( \frac{3}{2}\right)^{4u}\beta^u \left\{3 v +
9\beta(1-v)\right\},
\label{lambafI} \\
\label{lambagI} \frac{\tilde\Lambda_g}{\kappa_g} &=& \frac
{\zeta}{4}\left( \frac{2}{3}\right)^{4v} \beta^{-v} \left\{ 3 u -
9 \beta (1+u)\right\}.
\ea
Note that the corresponding
cosmological constants for the metrics (\ref{spheg}-\ref{sphef}) are not
determined solely by the vacuum energies $\rho_f$ and $\rho_g$.
They also contain a contribution from the interaction term in the
Lagrangian. This contribution depends not only on the parameters
$\zeta$ and $u$ (recall that $v=1/2 -u$), but also on the
arbitrary integration constant $\beta$, as seen from (\ref{lambafI}-\ref{lambagI}). \\

A crucial observation \cite{Blas:2005yk} is that even
without assuming any specific form for the functions $p(r)$ and $q(r)$, the Type I ansatz
leads to energy momentum tensors of the form (\ref{EQMMOT}). Here, we would like to clarify
the reason for that, and to show that Type I solutions (as well as some generalizations
thereof) exist also in the generic case (\ref{generi}).  Indeed, for arbitrary metrics $f$ and $g$
\ba
\label{emf}
f^{\m\a}T^f_{\a\n}=-2\zeta(g/f)^{u}\left(v V \delta^\m_\n-\sum_{n} n ({\mathcal M}^n)^\m_\n\ V^{(n)}\right),\\
\label{emg}
g^{\m\a}T^g_{\a\n}=-2\zeta(g/f)^{-v}\left(u V \delta^\m_\n+\sum_{n} n({\mathcal M}^n)^\m_\n\ V^{(n)}\right),
\ea
where we have introduced the notation
$$
V^{(n_1,...,n_l)}\equiv\frac{\pd^l V}{\pd \t_{n_1}\cdots\pd\t_{n_l}},
$$
where $l$ is the number of derivatives.
Moving to the frame where both metrics are diagonal,
the matrix ${\mathcal M}=f^{-1}\cdot g$ can be put to the diagonal form with eigenvalues $\l_i$.
Two arbitrary metrics $g_{\m\n}$
and $f_{\m\n}$ which are solutions of the vacuum Einstein equations will be solutions for bi-gravity
if all the $\t_n$ are constant and the eigenvalues of the matrix
\be
\sum n({\mathcal M})^\m_\n\ V^{(n)},
\ee
entering (\ref{emf}-\ref{emg}) are all equal to each other. Note that for a given ansatz,
the constancy of the traces (or of the eigenvalues) is a frame independent notion.
The equations of motion will be then satisfied provided that
\ba
\label{Lambdaf}
\Lambda_f=-2\kappa_f\zeta(g/f)^{u}\left(v V-\frac{1}{4}\sum_{n} n\t_n\ V^{(n)}\right)+\kappa_f\r_f,\\
\label{Lambdag}
\Lambda_g=-2\kappa_f\zeta(g/f)^{-v}\left(v V+\frac{1}{4}\sum_{n} n\t_n\ V^{(n)}\right)+\kappa_g\r_g,
\ea
where  $\Lambda_f$ and $\Lambda_g$ are the cosmological
constants.

Remarkably, the non-trivial background (\ref{spheg}-\ref{sphef}) has the property that the
eigenvalues of ${\mathcal M}$ are constant
$$
\l_i=\{\g^{-1},\g^{-1},\g^{-1},\b\g^{-1}\},
$$
which implies
$$
\t_n=\g^{-n}(3+\b^n), \quad \det[{\mathcal M}]=\b\g^{-4}.
$$
Thus, it is enough to impose
$$
\sum_{n} n({\mathcal M}^n)^\m_\n\ V^{(n)} \propto \delta^\m_\n.
$$
In the frame where ${\mathcal M}$
 is diagonal the previous
combination is a constant diagonal matrix with only two different constant eigenvalues
$$
\left\{\sum_n n\b^n\g^{-n} \ V^{(n)},
\sum_n n \g^{-n}\ V^{(n)}\right\}.
$$
Both eigenvalues will coincide when
\be
\label{typeIcondo}
\sum_n n \g^{-n}(-1+\b^n)\ V^{(n)}=0.
\ee
This tells us that for
any potential there will exist non-trivial solutions with certain $\g$ and $\b$ satisfying
(\ref{typeIcondo}) (note that the values of $V^{(n)}$ depend also on $\b$ and $\g$).
In addition, (\ref{Lambdaf}-\ref{Lambdag}) must be satisfied. These are three equations for the parameters
$\Lambda_f$, $\Lambda_g$, $\b$ and $\g$. Therefore, it is clear that
one of the effective cosmological constants can be chosen arbirarily.
It has the status of an integration constant.\\

Another interesting class of solutions is obtained by taking $f$ and $g$ proportional to each other,
but otherwise arbitrary
\be
\label{prop}
f_{\mu\nu}=\gamma(x) g_{\mu\nu}.
\ee
In this case, the matrix ${\mathcal M}$
is proportional to the identity ${\mathcal M}^\m_\n=\g^{-1}\delta^\m_\n$
and the energy-momentum
tensors (\ref{emf}-\ref{emg}) read
\ba
\label{Lambtilde}
\tilde \Lambda_f\delta^\m_\n\equiv \kappa_ff^{\m\a}T^f_{\a\n}=
-2\zeta \kappa_f\g^{-4u}\left(v V-\sum_n n\g^{-n}\  V^{(n)}\right)\delta^\m_\n\nonumber\\
\tilde \Lambda_g\delta^\m_\n\equiv \kappa_gg^{\m\a}T^g_{\a\n}=
-2\zeta \kappa_g\g^{4v}\left(u V+\sum_n n \g^{-n}\ V^{(n)}\right)\delta^\m_\n.
\ea
Thus, for any matter content
this term just adds to the vacuum energy. From Bianchi identities
$\tilde\L_f$ and $\tilde \L_g$ must be constant, and $f$ and $g$ must then be solutions of the
vacuum Einstein's equations. Generically, the expressions for $\tilde\Lambda_{f,g}$ depend on $\gamma$, so that
they imply a constant $\gamma$. In this case, the parameter $\gamma$
is determined through Einstein's equations by noting that (\ref{prop}) implies
\be
\label{Lambdas}
R_g=\gamma R_f.
\ee
Clearly, this class will include solutions in the Schwarzschild-(A)dS family, although non-spherically
symmetric solutions are possible as well. Note also that such solutions can easily be
generalized to multigravity theories by deconstructing 5D metrics with a
warp factor \cite{Deffayet:2003zm}.

Let us now turn to the study of perturbations to some of these background solutions.

\section{Perturbations around Lorentz-breaking biflat metrics}

In a theory with two metrics with Einstein-Hilbert kinetic terms and no
interaction, there are 4+4 ADM Lagrange multipliers.
When we add a non-derivative interaction which preserves diagonal diffeomorphisms, 4 combinations of these may in principle
appear non-linearly in the action \cite{Damour:2002ws}. For these, their equation of motion relates them to the other variables, but they do not
lead to further constraints. Thus, we have a minimum of 4 and a maximum of 8 Lagrange multipliers for 20 metric
components. Hence, we generically expect a maximum of $(10-4)+(10-8)=6+2=8$ degrees of freedom and a
minimum of $(10-8)\times 2=2+2=4$. In a Lorentz-invariant context, the first possibility corresponds to a massless and
a massive graviton, whereas the second would correspond to two massless gravitons. In the Lorentz breaking
context, it is possible to have a massive graviton with just two physical polarizations \cite{Dubovsky:2004ud,Gabadadze:2004iv}.\\

Let us consider a general potential $V[\{\t_n\}]$ as in (\ref{generi}).
As we showed in the last section, the vacuum energies $\r_f$ and $\r_g$ can be tuned so that the previous
 potential  has asymptotically bi-flat solutions.
At large distances from the origin, these take the form
\ba \label{BIMETMIN}
g_{\mu\nu}=\eta_{\mu\nu}, \quad f_{\mu\nu}=\gamma \tilde \eta_{\mu\nu},
\ea
where
 \be
\tilde{\eta}_{\mu\nu}=\eta_{\mu\nu}-\frac{\beta-1}{\beta}
\delta_\mu^0\delta_\nu^0,
\ee
and $\eta_{\mu \nu} = \mathrm{diag}(1,-1,-1,-1)$.
The parameters $\gamma$ and $\beta$ are related by Eq. (\ref{typeIcondo}).
For $\beta\neq 1$ \footnote{For $\beta =1$, we have proportional flat
metrics the perturbations of which can be obtained from the flat
space-time limit of the calculations done in the next section.},
we cannot simultaneously write both metrics in the canonical form $\eta_{\mu \nu}$, and Lorentz invariance
breaks down to spatial rotations.
It will be convenient to introduce the general perturbation in the form
\ba
f^{\mu\nu}&=&\gamma^{-1}\big(\tilde{\eta}^{\mu\nu}+h_{f}^{\phantom{f}\mu\nu}\big),\\
g_{\mu\nu}&=&\eta_{\mu\nu}+h^g_{\phantom{f}\mu\nu}, \ea
where $\tilde \eta^{\m\n}$ is the inverse of $\tilde \eta_{\m\n}$. The
perturbation to the metric $f$ has been defined with the upper indices,
just because this simplifies the manipulations which yield the action
quadratic in the perturbations shown below.
For the remainder of this section, all space-time indices will be
raised and lowered with the canonical Minkowski metric $\eta_{\mu \nu}$.
The interaction Lagrangian quadratic in perturbations then reads
\ba
\label{lagran}
\tilde L_{int}&\equiv& L_{int}-\sqrt{-g}\r_g-\sqrt{-f}\r_f\nonumber=\\
&&-\frac{M^4}{8}\Big\{
n_2(h^g_{\phantom{f}ij}+h_f^{\phantom{f}ij})(h^g_{\phantom{f}ij}+h_f^{\phantom{f}ij})
+n_0 (h^g_{\phantom{f}00}+\beta^{-1}h_f^{\phantom{f}00})(h^g_{\phantom{f}00}+\beta^{-1}
h_f^{\phantom{f}00})\nonumber\\
&&
-2n_4(h^g_{\phantom{f}00}+\beta^{-1}h_f^{\phantom{f}00})(h^g_{\phantom{f}ii}
+h_f^{\phantom{f}ii})+n_3 (h^g_{\phantom{f}ii}+h_f^{\phantom{f}ii})^2\Big\},\label{lame}
\ea
where, after imposing (\ref{typeIcondo})
\ba
M^4&=&4\zeta\left(\frac{\g^4}{\b}\right)^v, \quad
n_0=3n_3-2n_4+n_2+\g\frac{\pd}{\pd \g}\left(\sum_n n \g^{-n}(-1+\b^n)V_0^{(n)}\right),\nonumber\\
n_2&=&-\sum n^2\g^{-n}V_0^{(n)},\quad
n_3=
u v V_0+\sum_n n[v-u] \g^{-n}V^{(n)}_0
-\sum_{m,n} n m \g^{-(n+m)}V_0^{(n,m)},\nonumber\\
n_4&=&n_0+\b \frac{\pd}{\pd \b}\left(\sum_n n \g^{-n}(-1+\b^n)V_0^{(n)}\right).\label{coefftypeI}
\ea
For the sake of simplicity, we will restrict to potentials $V[\{\t_n\}]$
for which  Eq. (\ref{typeIcondo}) is independent\footnote{
The case where (\ref{typeIcondo}) is satisfied independently of $\b$ or $\g$ leads to the condition
\be
\label{noNcorr}
3n_3-3n_0+n_2=0, \quad n_4=n_0,
\ee
which, as we shall see, corresponds to the case of no corrections to the Newton's law. An example of
an interaction where these conditions are satisfied is a potential which is only a function of the
ratio of determinants of $f$ and $g$; that is $V\left[\{\tau_n\}\right] = V[f/g]$. In this particular case,
there is an enhanced symmetry under independent ``non-diagonal" unimodular diffeomorphisms, which do not change
the value of the determinants of the respective metrics.}
of $\beta$, and determines $\gamma$. From equation (\ref{coefftypeI}), this implies $n_0=n_4$.
 In particular, this class includes
the interaction (\ref{interaction}), which, as we shall see,
leads to a rather pathological behaviour for the perturbations. On the other hand, it is general enough to
be representative of generic choices of potentials.

In Refs. \cite{Rubakov:2004eb,Dubovsky:2004sg} the case of a single graviton with a Lorentz violating mass term has been discussed.
For comparison with those references, it will be useful to introduce
$$m_0^2=-c n_0, \ m_1^2=0, \ m_2^2=c n_2, \ m_3^2=-c n_3, \ m_4^2=-c n_4,$$
where $c>0$ is an irrelevant constant which has the dimensions of mass squared.

Note that the components $h^g_{\phantom{f}0i}$ and $h_f^{\phantom{f}0i}$ are absent from (\ref{lame}).
As noted in \cite{Berezhiani:2007zf} the absence of such terms is a consequence of invariance under
diagonal diffeomorphisms in this background.
In the case of a single graviton (with a Fierz-Pauli kinetic term), the absence of $h_{0i}$ in the mass term leads to a very
interesting behaviour \cite{Dubovsky:2004sg,Dubovsky:2004ud,Dubovsky:2005dw}, where the two polarizations of the massless
graviton acquire mass, while all the other modes
 do not propagate.\footnote{It should be stressed that
the absence of $0i$ components is a peculiarity of the background considered. By suitable adjustment of the vacuum energies,
the theory we are considering also
admits the Lorentz preserving vacuum of type II, where $f_{\mu\nu}=g_{\mu\nu}=\eta_{\mu\nu}$. In that case, the interaction term
leads to the Fierz-Pauli mass term for a combination of the two gravitons. This mass term does contain the $0i$ components.}

Let us now investigate whether a similar phenomenon occurs in our model. The situation is not directly reducible to
that of a single graviton, since the equations of motion are not diagonal. Also, the kinetic term breaks the Lorentz invariance.
It is convenient to decompose the perturbations into
irreducible representations
of the spatial rotations,
\ba
\label{decompAph}
 h^X_{\phantom{f}00}&=&2 A^X\nonumber,\\
 h^X_{\phantom{f}0i}&=& B^X_{,i}+V^X_i\nonumber,\\
 h^X_{\phantom{f}ij}&=&2\psi^X\delta_{ij}-2E^X_{,ij}-2F^X_{(i,j)}-t_{ij}^X,\label{irrep}
\ea
where $t^X_{\phantom{X}ii}= t^X_{\phantom{X}ij,i}= V^X_{\phantom{X}i,i}=
F^X_{\phantom{X}i,i}=0$ for $X=f,g$, and all space-time indices are raised and lowered with
the metric $\eta_{\mu\nu}$.\\

To second order in the perturbations, the kinetic terms in (\ref{action}) can be written in terms of these
scalar, vector and tensor variables as:
\ba
L_K&=&\frac{1}{2\kappa_g}\Big\{-\frac{1}{4}t^g_{ij}\Box t^g_{ij}
-\frac{1}{2}\left(V_i^g+\dot F_i^g\right)\Delta
\left(V^g_i+\dot F^g_i\right)+4\Delta \psi^g \left(A^g-\dot B^g-\ddot E^g\right)\nonumber \\
&&-2\psi^g \Delta \psi^g-6 (\dot\psi^g)^2\Big\}
+\frac{1}{2\tilde \kappa_f}\Big\{-\frac{1}{4}t^f_{ij}\tilde \Box t^f_{ij}
-\frac{\beta^{-1}}{2}\left(V^f_i+\b\dot F^f_i\right)\Delta
\left(V^f_i+\b\dot F^f_i\right)\nonumber \\
&&\hspace{1cm}+ 4\beta^{-1} \Delta \psi^f \left(A^f-\b\dot B^f-\b^2\ddot E^f\right)
-2 \psi^f \Delta \psi^f-6 \beta(\dot \psi^f)^2\Big\}
\ea
where $\tilde \Box=\tilde \eta^{\m\n}\partial_\m \partial_\n$,
 $\tilde \kappa_f=\gamma^{-1} \beta^{1/2}\kappa_f$ and dot means a derivative with respect to time.
At the linear level, the transformations generated by independent
diffeomorphisms $\delta x^\mu=\xi^\mu_X$ in each one of the metrics can be expressed as
\ba
\delta h^g_{\phantom{g}\m\n}=2\partial_{(\m}\xi^g_{\n)},
\quad \delta h^f_{\phantom{g}\m\n}=2\eta_{\b(\m|}\tilde \eta^{\a\b}\partial_\a \xi^f_{|\n)}.
\ea
Note that the kinetic term is written in terms of the following quantities:
\ba
t_{ij}^g, \ V_i^g+\dot F_i^g, \ \psi,\ A^g-\dot B^g-\ddot E^g, \nonumber\\
t_{ij}^f, \ V_i^f+\b\dot F_i^f, \ \psi,\ A^f-\b\dot B^f-\b^2\ddot E^f,
\ea
which are invariant under both gauge symmetries.
On the other hand, the full action (including the mass terms),
is invariant only under the diagonal gauge symmetry
\be
\xi_\m^g=\xi^f_\m.
\ee
No second order scalar combination of $h^X_{\phantom{X}0i}$ is invariant under this
gauge symmetry, which implies that those terms are always absent (cfr. (\ref{lame})).
We may now analyze the propagating degrees of freedom.

\subsection{Tensor Modes}

The linearized Lagrangian for the tensor and vector modes can be expressed as
\ba
\label{tensandvec}
 L_{t,v}&=&\frac{1}{2\kappa_g}\Big\{-\frac{1}{4}t^g_{ij}\Box t^g_{ij}
-\frac{1}{2}\left(V_i^g+\dot F_i^g\right)\Delta
\left(V^g_i+\dot F^g_i\right)\Big\}\nonumber\\
&&+\frac{1}{2\kf}\Big\{-\frac{1}{4}t^f_{ij}\tilde \Box t^f_{ij}
-\frac{\beta^{-1}}{2}\left(V^f_i+\b\dot F^f_i\right)\Delta
\left(V^f_i+\b\dot F^f_i\right)\Big\}\nonumber\\
&&-\frac{M^4}{8}\Big\{
n_2(t_{ij}^g+t_{ij}^f)^2-2n_2(F_i^g+F_i^f)\Delta(F_i^g+F_i^f)\Big\},
\ea
where $\tilde\kappa_f=\gamma^{-1} \beta^{1/2}\kappa_f$.
The corresponding equations of motion in Fourier space read
\ba \label{DIS1}
\omega^2 t_{ij}^g&=&{\bf k}^2t_{ij}^g+
\kappa_g M^4n_2(t_{ij}^g+t_{ij}^f)\\ \label{DIS2}
\beta\omega^2 t_{ij}^f&=&{\bf k}^2t_{ij}^f+
\kf M^4 n_2(t_{ij}^g+t_{ij}^f)
\ea
from which we obtain the dispersion relations
\be
\label{dispete}
\omega^2_\pm=\frac{1}{2\beta}\Big((\beta+1){\bf  k}^2+ \kappa_0 M^4\pm\sqrt{
((\beta+1){\bf k}^2+ \kappa_0 M^4)^2-4\b {\bf k}^2(\kappa_1 M^4+{\bf k}^2)}\Big)
\ee
where  $\kappa_0=n_2  (\b\kappa_g+\kf)$ and $\kappa_1= n_2  (\kappa_g+\kf)$.

At high energies, we have
\be
\omega^2_+\approx{\bf k}^2,\quad \omega^2_-\approx{\beta}^{-1 }{\bf k}^2.
\ee
In this limit, each one of the two gravitons propagates in its own metric
 (with the corresponding ``speed of light" \footnote{Superluminal propagation has
 previously been considered in several contexts (see e.g. \cite{Babichev:2006vx} for a
 recent discussion). Clearly, such propagation cannot by
 itself be considered pathological. Indeed, in the present case
 we always have superluminal propagation from the point of view of one of the metrics,
 whereas there is not any superluminal propagation from the point of view of the other
 metric. Nevertheless, the global structure of non-linear bi-gravity solutions is complicated in
 general, and its interpretation is far from trivial \cite{Blas:2005yk,Blas:2005P}.})
along null directions $k^{\mu}=(\omega,{\bf k})$ satisfying
$$g^{\m\n}_X k_\m k_\n \approx 0.$$
The low energy expansion of (\ref{dispete}) is given by
\ba
\omega^2_-&=&\frac{\kappa_1}{\kappa_0} {\bf k}^2 +O({\bf k}^4),\\
\label{Ommexp}
\omega^2_+&=&\frac{\kappa_0 M^4}{\beta}+
\left({\tilde \kappa_f+\beta^2\kappa_g \over \beta\tilde\kappa_f+
\beta^2\kappa_g}\right){\bf k}^2+ O({\bf k}^4).\label{masgru}
\ea
The first dispersion relation corresponds to two massless polarizations
 which propagate at the ``intermediate" speed
$$
c_s^2 = {\omega_-^2\over {\bf k}^2}= {\kappa_1 \over \kappa_0} = {\kappa_g +\tilde\kappa_f \over \beta \kappa_g +\tilde \kappa_f}.
$$
Note that for $\beta>1$ we have $\beta^{-1}<c_s^2<1$, while for $\beta<1$ we have $1<c_s^2 <\beta^{-1}$.
The second dispersion relation, Eq. (\ref{masgru}), corresponds to two massive polarizations.
It is easy to check that the graviton polarizations are stable and tachyon free as long as $\kappa_0>0$, in the whole range of
momenta ${\bf k}$. The second dispersion relation (\ref{masgru}) corresponds to the massive graviton.

\subsection{Vector Modes}

From the Lagrangian (\ref{tensandvec}), we find that $V_i^g$ and $V_i^f$ do not appear in the interaction term.
Varying with respect to the vector fields we have,
\ba
\label{vector}
\Delta(V_i^g+\dot{F}_i^g)&=& 0 \label{tor1}\\
\Delta \left(\dot{V}^g_i+\ddot{F}^g_i\right)  &=& -M^4n_2\kappa_g \Delta \left(F^g_{i} +F^f_{i} \right)\label{tor2} \\
\Delta(V_i^f+\beta \dot{F}_i^f) &=& 0\label{tor3}\\
\Delta \left(\dot{V}^f_i+\beta\ddot{F}^f_i\right)  &=& - M^4 n_2\kf \Delta \left(F^g_{i} +F^f_{i} \right).
\ea
We can always use the diagonal diffeomorphism invariance to work in
the gauge where $V_i^g=0$. It then follows from (\ref{tor1}) that $F_i^g=F_i(\vec x)+f_i^g(t)$, where $F_i$ are arbitrary
functions of position and $f_i$ are arbitrary functions of time. The latter are in fact
irrelevant, because $F_i^X$ enters the metric only through spatial derivatives. Formally, we may describe this as a gauge
symmetry $F_i^X\to F_i^X + f_i^X(t)$, which we can use in order to write, without loss of generality,
$$F_i^g=F_i(\vec x).$$ It then follows from
(\ref{tor2}) that $$F_i^f= - F_i(\vec x),$$ where again we eliminate the additive time dependent part. Finally, from (\ref{tor3})
we obtain $$V_i^f=\tilde f_i(t),$$
where $\tilde f_i$ are new arbitrary functions of time. This is not a desirable situation, since it means that the initial
conditions do not determine the future evolution of $V_i^f$. Technically, the absence of the fields $V_i^g$ and $V_i^f$ in the mass
term leads an enhanced gauge symmetry in the {\em linearized} Lagrangian. Indeed, we can consider independent
gauge transformations for each of the metrics
\be
h_{\m\n}\mapsto h_{\m\n}+2\partial_{(\m}\xi^h_{\n)},\quad
l_{\m\n}\mapsto l_{\m\n}+2\partial_{(\m}\xi^l_{\n)},
\ee
of the form $\xi_i^X=\xi_i^X(t)$. As we have discussed, these do not affect the $F_i^X$, but can be used to give both of the $V_i^X$
an arbitrary time dependence.

\subsection{Scalar Modes}

The Lagrangian for the scalar modes can be expressed as
\ba
&&L_s=\frac{1}{\kappa_g}\Big\{2\Delta \psi^g \left(A^g-\dot B^g-\ddot E^g\right)-\psi^g \Delta \psi^g-3 (\dot\psi^g)^2\Big\}
+\frac{1}{\kf}\Big\{
 2\beta^{-1} \Delta \psi^f \left(A^f-\b\dot B^f-\b^2\ddot E^f\right)\nonumber \\
&&
-\psi^f \Delta \psi^f-3 \beta(\dot \psi^f)^2\Big\}-\frac{M^4}{2}\Big\{
n_2\{3(\psi^g+\psi^f)^2+(\Delta(E^g+E^f))^2-2(\psi^g+\psi^f)\Delta (E^g+E^f)\}\nonumber\\
&&+n_0 \{(A^g+\beta^{-1}A^f)\left(A^g+\beta^{-1}A^f-2[3(\psi^g+\psi^f)-\Delta(E^g+E^f)]\right)\}+n_3 \{3(\psi^g+\psi^f)-\Delta(E^g+E^f)\}^2\Big\}.\nonumber
\ea
Let us first study the non-homogeneous modes.
The mass terms do not depend on $B^g$ nor on $B^f$, so those fields are Lagrange multipliers, just as in Einstein's gravity.
Variation with respect to these fields yields
\be
\Delta \dot \psi^g=\Delta \dot \psi^f=0.
\ee

The variation with respect to $A^g$ and $A^f$ yields the constraints
\ba
A^g&=&-\b^{-1} A^f+3(\psi^g+\psi^f)-\Delta(E^g+E^f)+\frac{2}{M^4 n_0\kappa_g}\Delta \psi^g, \nonumber\\
\label{Af}
\psi^g&=&
\frac{\kappa_g}{\kf} \psi^f+f(t).
\ea
Once we substitute the first of these constraints in the Lagrangian, the quadratic term in $E^h$ and
$E^l$ takes the form
\be
(n_2-n_0+n_3)(E^h+E^l)^2.
\ee
We can now distinguish two different cases, neither of them with propagating
scalar degrees of freedom. First, if the coefficient $n_2-n_0+n_3$ does not cancel,
the equations of motion for $E^h$ and $E^l$ result in a new constraint which determines these fields,
and upon substitution into the Lagrangian we are left without any scalar degrees of freedom.
If the coefficient cancels, as happens for the potential (\ref{interaction}),
 $E^g$ and $E^f$ are Lagrange multipliers appearing
in the gauge invariant combination $E^h+E^l$. After using (\ref{Af}), the variation with respect to $E^h$ yields
\be
\Delta \psi^g=\Delta \psi^f=0.
\ee
The Lagrangian cancels after substitution of these constraints, and there are no propagating degrees
of freedom. Note that in this last case the combination $E^h+E^l$, is not determined by the equations of motion.
Again, this is not a desirable feature, since it means that the value of this combination, which is gauge invariant under
the diagonal diffeomorphisms, is not predicted by the linear theory. Nevertheless, we expect that higher order
terms in the expansion will determine $E^h+E^l$, since there is no symmetry in the non-linear Lagrangian under which this quantity
can be ``gauged" to arbitrary spacetime dependence.

Concerning the homogeneous modes, after using the constraints we are left with two modes $\psi^{f}$ and $\psi^g$ which have
a negative definite kinetic term. Nevertheless, the dispersion relations for the degrees of freedom which
diagonalize the equations of motion are $\omega^2=0$ and $\omega^2= M^4 n_2(\tilde \kappa_f +\kappa_g)>0$, so there
is no classical instability associated to these modes.

\subsection{Coupling to Matter and vDVZ discontinuity}

All known explicit and non-singular
exact solutions of bigravity (which we reviewed in Section I) are also solutions of GR. This immediately suggests that
the vDVZ discontinuity may be absent altogether in this theory at the non-linear level. Also, from the analysis of perturbations
done in the previous Section around the Lorentz breaking background, it is clear that the situation here is very different from
that of ordinary massive gravity. The massive spin-2 graviton has only two physical polarizations (as opposed to the five
polarizations of the ordinary FP massive graviton), and there are no propagating vectors or scalars.

Let us consider the coupling of the linearized theory to conserved sources. To this end, we introduce the couplings
\be
S_{matt}=\frac{1}{4} \int \di^4 x \left(\lambda_g h^g_{\phantom{g}\m\n} T_g^{\m\n}+\lambda_f h^f_{\phantom{g}\m\n} T_f^{\m\n}\right),
\ee
where $T_g^{\m\n}$ and $T^f_{\m\n}$ are conserved, i.e. $\partial_{\m}T_g^{\m\n}=0$ and
$\eta_{\r\m}\tilde \eta^{\r\a}\partial_\a T_f^{\m\n}=0$. In terms of the decomposition (\ref{irrep}), we have
\ba
S_{matt}&=&\frac{\lambda_g}{4}\int \di^4 x \left(-t_{ij}^g T_g^{ij}+2 T_g^{0i}(V^g_i+\dot F^g_i)+2 T_g^{00}\Phi^g+2T_g^{ii}\psi^g\right)
\nonumber\\
&+&\frac{\lambda_f}{4}\int \di^4 x \left(-t_{ij}^f T_f^{ij}+2 T_f^{0i}(V^f_i+\b\dot F^f_i)+2 T_f^{00}\Phi^f+2T_f^{ii}\psi^f\right).
\ea
where we have introduced the gauge invariant combinations
$$\Phi^g\equiv A^g-\dot B^g-\ddot E^g, \quad \quad \Phi^f\equiv
A^f-\b\dot B^f-\b^2\ddot E^f.$$
Inverting the equations of motion for the tensor modes in the presence of the source $T^{ij}$, we find
\be
t_{ij}^g=\frac{\l_g ({\bf k}^2-\b \omega^2+\kf M^4n_2)T^g_{ij}-\l_f \kappa_g M^4 n_2 T_{ij}^f}{\omega^2\{\b \omega^2-
(\kf+\b\kappa_g)M^4n_2\}+{\bf k}^2\{
(\kf+\kappa_g)M^4n_2-(\b+1) \omega^2\}+{\bf k}^4},
\ee
and an analogous expression for $t_{ij}^f$:
\be
t_{ij}^f=\frac{\l_f ({\bf k}^2- \omega^2+\kappa_g M^4n_2)T^f_{ij}-\l_g \tilde\kappa_f M^4 n_2 T_{ij}^g}{\omega^2\{\b \omega^2-
(\kf+\b\kappa_g)M^4n_2\}+{\bf k}^2\{
(\kf+\kappa_g)M^4n_2-(\b+1) \omega^2\}+{\bf k}^4}.
\ee
In the limit $M^4\rightarrow 0$ this reduces to the standard expression for linearized GR.

For the vector modes, the equations of motion read
\ba
\Delta(V_i^g+\dot F^g_i)&=&\lambda_g \kappa_g T^{0i}_g  \nonumber \\
\Delta \left(\dot{V}^g_i+\ddot{F}^g_i\right)  &=& -M^4n_2\kappa_g \Delta\left(F^g_{i} +F^f_{i} \right)+\lambda_g \kappa_g\dot
T^{0i}_g \\
\Delta(V_i^f+\b\dot F^f_i)&=&\lambda_f \b\kf T^{0i}_f  \nonumber \\
\Delta \left(\dot{V}^f_i+\b\ddot{F}^f_i\right)  &=& -\kf \b^{-1}M^4n_2 \Delta\left(F^g_{i} +F^f_{i} \right)
+\lambda_f \kf\b\dot
T^{0i}_f.
\ea
It follows immediately that $\Delta(F_i^g+F_i^f)=0$, and therefore the term proportional to $M^4$ vanishes.
This means that there is no difference with the GR results for each one of the metrics.

For the scalar part, we may start with variation with respect to $B_i^X$, which yields the constraints
\be
\dot C_X = 0
\label{consu}
\ee
where
$$
C_g\equiv 4\Delta \psi^g +\lambda_g\kappa_g T_g^{00}, \quad\quad C_f\equiv 4\Delta \psi^f +\lambda_f\tilde\kappa_f\beta T_f^{00}.
$$
Variation with respect to $A^X$ gives
\be
C_f = C_g
\ee
and
\be
C_+\equiv C_f + C_g = 2 M^4(\tilde\kappa_f + \kappa_g )(A_+ -3 \psi_+ + \Delta E_+) n_0,\label{nnn}
\ee
where $A_+= A_g + \beta^{-1} A_f$, $\psi_+ = \psi_f + \psi_g$, and $E_+ = E_f + E_g$.
Variation with respect to $\Delta E^X$ yields, with the help of (\ref{consu}),
\be
n_0 A_+ = (n_2 + 3n_3) \psi_+ - (n_2 + n_3) \Delta E_+. \label{cep}
\ee
Substituting into (\ref{nnn}), we have
$$
C_+ = 2 M^4(\tilde\kappa_f + \kappa_g )[(n_2+3n_3-3n_0)\psi_+ - (n_2+n_3-n_0) \Delta E_+]
$$
and using (\ref{consu}), we have
\be
4(n_2 - n_0 + n_ 3) \Delta^2 \dot E_+ = - (n_2+3 n_3 -3 n_0)
(\lambda_f\tilde\kappa_f \beta\dot T_f^{00}+\lambda_g\kappa_g \dot T_g^{00}). \label{consu2}
\ee
For $(n_2-n_0+n_3)\neq 0$, this determines $\dot E_+$ in terms of the sources.
The solution will depend on an arbitrary time independent mode $E_0(x)$.

For the singular case $(n_2-n_0+n_3)=0$, Eq. (\ref{consu2}) do not determine
$E_+$ at all. Instead, it imposes some non-trivial equations to be satisfied by the sources,
\be
\lambda_f\tilde\kappa_f \beta\dot T_f^{00}=-\lambda_g\kappa_g \dot T_g^{00} \label{urest}
\ee
which seem hard to motivate. Thus, coupling to the sources seems rather inconsistent in this case,
unless $(n_2+3 n_3-3n_0)=0$ as well. But this would imply $n_2=0$, in which case the tensor modes are
massless.

In the generic case, the solution for the $\psi$ potentials is of the form
\ba
\Delta \psi^g = -{\kappa_g \lambda_g \over 4} T^{00}_g + {1\over 8} C_+ (\vec x),\nonumber\\
\Delta \psi^f = -{\tilde\kappa_f \lambda_f \beta \over 4} T^{00}_f + {1\over 8} C_+ (\vec x).\label{mosca}
\ea
where $C_+(\vec x)$ is entirely determined by initial conditions.

Finally, variation with respect to $\psi_f$ and $\psi_g$ leads [after use of (\ref{mosca})]
to the following equations for the gauge invariant
potentials:
\ba
\Delta \Phi^g&=&-\frac{\kappa_g\lambda_g}{4}\left( T^{00}_g+T^{ii}_g-{3\over \Delta} \ddot T^{00}_g\right)+ {1\over 8} C_+
+\kappa_g M^4 n_2 \Delta E_+ ,\\
\beta^{-1}\Delta \Phi^f&=&-\frac{\tilde\kappa_f\lambda_f}{4}\left( \beta T^{00}_f+T^{ii}_f-{3\over \Delta} \beta^2 \ddot T^{00}_f\right)
+{1\over 8} C_+
+\tilde \kappa_f M^4 n_2 \Delta E_+.\label{mosco}
\ea
where
\be
\Delta E_+= - {1\over n_2 +n_3 -n_0} \left[ {n_2 + 3 n_3 -3 n_0 \over 4 \Delta} \left(
\kappa_g\lambda_g T^{00}_g + \tilde\kappa_f\lambda_f \beta T^{00}_f -
C_+\right) + {1\over 2 M^4 (\tilde\kappa_f + \kappa_g)} C_+\right].
\label{ep}
\ee
In general, the solution depends on an arbitrary ``initial" function $C_+(\vec x)$.
This corresponds to a mode with dispersion relation $\omega^2=0$ in the linear theory.
It was argued in \cite{Dubovsky:2004sg} that in such cases, from higher order terms the expected dispersion
relation will be of the form $\omega^2\sim p^4$, and in this sense $C_+$ corresponds to
a slowly varying ``ghost condensate" \cite{Arkani-Hamed:2003uy}. In what follows, we shall take the
initial condition $C_+(\vec x)=0$.

For $n_2-n_0+n_3 \neq 0$, the solution is of the form
\be
\Delta \psi^g = -{\kappa_g \lambda_g \over 4} T^{00}_g,\quad
\Delta \psi^f = -{\tilde\kappa_f \lambda_f \beta \over 4} T^{00}_f,
\ee
and
\ba
\Delta \Phi^g=&-&\frac{\kappa_g\lambda_g}{4}\left( T^{00}_g+T^{ii}_g-{3\over \Delta} \ddot T^{00}_g\right)
-\left({\kappa_g M^4 n_2\over 4\Delta}\right) {n_2 + 3 n_3 -3 n_0 \over n_2 +n_3 -n_0} \left(
\kappa_g\lambda_g T^{00}_g + \tilde\kappa_f\lambda_f \beta T^{00}_f \right)
,\\
\Delta \Phi^f=&-&\frac{\tilde\kappa_f\lambda_f\beta}{4}\left( \beta
T^{00}_f+T^{ii}_f-{3\over \Delta} \beta^2 \ddot T^{00}_f\right)
-\left({\tilde\kappa_f\beta M^4 n_2\over 4\Delta}\right) {n_2 + 3
n_3 -3 n_0 \over n_2 +n_3 -n_0} \left( \kappa_g\lambda_g T^{00}_g +
\tilde\kappa_f\lambda_f \beta T^{00}_f \right).\nonumber
\label{mosco} \ea Hence, there is a well behaved massless limit,
with corrections of order $M^4\Delta^{-2}$ to the gauge invariant
potentials $\Phi$ and $\psi$. This means, in particular, that there
is no vDVZ discontinuity. This is quite analogous to the ``half
massive gravity" model discussed by Gabadadze and Grisa
\cite{Gabadadze:2004iv} (see also \cite{Dubovsky:2004ud}). The additional terms lead to corrections to the Newtonian
potential. The sign of this correction can be positive or negative, depending on the values of
the numerical coefficients $n_i$.
For isolated sources, such corrections scale like the square of the graviton mass $m^2 \sim \kappa M^4$
times the ``Schwarzschild" radius $r_s$ corresponding to the given source, and grow linearly
with the distance $r$.
Parametrically, the potential takes the form
$$\Phi \sim \phi_N + m^2 r_s r,$$
where $\phi_N$ is the standard
Newtonian potential. Linear theory breaks down at large distances, when the second term is of order unity.
It would be interesting to try and match this solution to a non-perturbative exact solution which is well behaved
at infinity.

As we stated before, the case of no correction
to the Newton's law corresponds to the case where (\ref{typeIcondo}) is independent of $\b$ or $\g$ (cfr.
(\ref{noNcorr})).\\

Finally, we note that the simple interaction term (\ref{interaction}) first considered in \cite{Isham:gm,Isham:1977rj}
happens to land on the special case $$n_2-n_0+n_3=0,$$
where the above expressions for the gauge invariant potentials are singular. The origin of the singularity is the following.
After substitution of the constraints (\ref{cep}), the linearized action no longer
depends on $\Delta E_+$. In particular, the
absence of
this variable results in the unwanted restriction (\ref{urest}) on the sources\footnote{
This accidental symmetry is similar to that which exists in ordinary massive gravity
where the linear action propagates 5 dof whereas a new ghost-like dof appears at the
non-linear level \cite{Deffayet:2005ys,Boulware:1973my}.
 However, in that case the accidental symmetry corresponds
to a symmetry of the massless theory and no further constraints are needed in the sources.}.
Nevertheless, beyond the linear order, the action will
depend on $\Delta E_+$, and hence the ``restriction" will no longer exist.
Rather, a nonlinear equation will determine the value of $\Delta E_+$.
Can we nevertheless try to find classical solutions in a perturbative expansion? The above considerations
suggest an expansion scheme for the singular case $n_2-n_0+n_3= 0$, where $E_+$
is treated as a much bigger quantity than the rest of the linearized fields\footnote{Some of the
linearized fields will be of the order of $E$ as is clear from (\ref{cep}).} (such as $\psi$).
Heuristically, the size of $\Delta E_+$ can be estimated as follows.
Instead of perturbing the flat solution Eq. (\ref{BIMETMIN}), we may consider the quadratic
action for perturbations around a solution
which differs from the original by $O(h)$. The expansion around this new solution will have\footnote{
All the coefficients will have corrections of order $O(h)$. However, for the
rest of coefficients one expects that they will yield second order small corrections.
}
 $$n_2-n_0+n_3=O(h).$$
From (\ref{mosca}), we have
$$
\Delta\psi \sim \kappa T \equiv\Delta \phi_N,
$$
where $\phi_N$ stands for the potential corresponding to the given source in Newton's theory.
From (\ref{mosco}), $\Delta\Phi \sim O(\kappa T) + O(m^2 \Delta E)$,
where $m^2\sim \kappa M^4$
denotes the graviton mass squared.
From
(\ref{ep}), we have  $\Delta E \times O(h) \sim O( \kappa T/\Delta )$.
This suggests the hierarchy
$$
\Delta E \gg \psi, \quad
\Phi\sim \max (\psi, m^2 E).
$$
Taking
 $n_2-n_0+n_3 \sim \max (\Phi,\Delta E)\sim \max (\psi, m^2 E,\Delta E)\sim \Delta E\left(
1+m^2/\Delta\right)$, this leads to the estimate
$$
(\Delta E)^2 \sim {\psi \over 1+ m^2/\Delta}.
$$
For distances shorter than the inverse graviton mass, we have
$
\Delta E \sim \phi_N^{1/2},
$
and hence we may expect
$$
\Phi \sim \phi_N + (m^2/\Delta) \phi_N^{1/2}.\quad\quad (\Delta \gg m^2)
$$
At distances which are large compared with the inverse graviton mass,
$\Delta E \sim (\Delta \phi_N /m^2)^{1/2}$, and we expect
$$
\Phi \sim (m^2/\Delta)^{1/2} \phi_N^{1/2}.\quad\quad (\Delta \ll m^2)
$$
These very crude arguments seem to indicate that, also in this special case,
there is no vDVZ discontinuity. However, for finite $m$,
there are significant modifications to the value of the ``gauge invariant" potential $\Phi$ which determines the motion of
slowly moving particles. For isolated sources, such modifications scale like $r_s^{1/2}$, where
$r_s$ is the ``Schwarzschild" radius corresponding to the given source. They grow with the distance as
$r^{3/2}$ below the graviton Compton wavelength $m^{-1}$, and as $r^{1/2}$ for larger distances. The potential $\Phi$
becomes of order one for $r\gtrsim m^{-2} r_s^{-1}$, beyond which we enter a non-perturbative regime. It would be
interesting to confirm this heuristic analysis in a numerical study of a spherically symmetric solution with sources.
This is left for further research.

\section{Perturbation theory of Proportional  de Sitter Metrics}

As stated in Section 2, another interesting class of solutions of bigravity can be
constructed from two proportional metrics with
a constant proportionality factor. Let us define our perturbations as
\ba
g_{\mu\nu}&=&\Omega_{\mu\nu}+h^g_{\mu\nu},\\
f^{\mu\nu}&=&\gamma^{-1}(\Omega^{\mu\nu}+h_f^{\mu\nu}).
\ea
All indices will be handled with the $\Omega_{\mu\nu}$ metric. \\

We first focus in the interaction term for a general potential (\ref{generi}).
Using (\ref{Lambtilde}) we can write
\ba
\label{lgenerbi}
\tilde L_{int}&=&\zeta (-g)^u(-f)^v V[\{\t_n\}] +
\sqrt{-g} \frac{\tilde \Lambda_g}{\kappa_g}+ \sqrt{-f}
\frac{\tilde \Lambda_f}{\kappa_f}\nonumber\\
&&=-\frac{1}{8\kappa_+}\sqrt{-\Omega}
\left\{m^2_t(h_g^{\m\n}+h_f^{\m\n})(h^g_{\m\n}+h^f_{\m\n})-m^2_s(h^g+h_f)^2\right\},
\ea
where indices are manipulated with the metric $\Omega_{\m\n}$, e.g. $h^g=\Omega^{\m\n}h^g_{\m\n}$,
 and
\ba
m^2_s&=&4\kappa_+\zeta \g^{4v}\left(-uvV_0
+(u-v)\sum_n n\g^{-n} V^{(n)}+
\sum_{n,m}  nm\g^{-(n+m)}V^{(n,m)}\right),\nonumber\\
m^2_t&=&-4\kappa_+\zeta \g^{4v}\sum_n n^2 \g^{-n}V^{(n)}.
\ea%
We have also introduced an
effective Newtons's constant $\kappa_+$ for later convenience.

Note that the massive graviton corresponds to
$h^+_{\mu\nu}=(h_g+h_f)_{\mu\nu}$. This is to be expected,
as for $h^g_{\mu\nu}=-h^f_{\mu\nu}$ the metrics are still
proportional and therefore the perturbations are standard massless gravitons of GR in vacuum.
Also, in the present set-up, $h^+_{\mu\nu}$ are the quantities invariant
under the diagonal diffeomorphisms.
Notice also that the mass term does not have in general a Pauli-Fierz form,
\be
m^2( h_+^2- h_+^{\m\n} h^+_{\m\n}).
\ee
This particular form can only be achieved by properly tuning the parameters.
This is in contrast with other ways of getting massive gravitons, such as dimensional reduction, where
the original symmetry group is much larger. Here, the degrees of freedom of the original theory are $8$ which
can be split into a massless graviton with $2$ polarizations and a massive
graviton with $6$ polarizations\footnote{
The number of degrees of freedom coincides with that of higher derivative gravity \cite{Stelle:1977ry}.}.
The expression of the massless graviton as a linear combination of the metric perturbations
will be given below.

From Eq. (\ref{lgenerbi})
we note that whenever $m_t=0$ there is an enhancement of the gauge symmetry, which
now admits all transformations which leave the traces $h_g$ and $h_f$ invariant \footnote{This
happens in the case when the derivative of Eq. (\ref{typeIcondo}) with respect to $\beta$
vanishes at $\beta=1$. For the case (\ref{interaction}) this amounts to $\g=2/3$.}.
This corresponds to the transverse subgroup of the diffeomorphisms, which
has been recently considered in \cite{Alvarez:2006uu}.  In this special case the gauge
symmetry is enough to have just two massless gravitons propagating
\footnote{At first sight, this seems to contradict the
results of Ref. \cite{Boulanger:2000rq}, where it is shown that we cannot have two massless interacting
gravitons. However, the starting point in \cite{Boulanger:2000rq} is a free Lagrangian
invariant under linearized diffeomorphisms. As shown in \cite{Alvarez:2006uu},
there are Lagrangians invariant under transverse diffeomorphisms which
propagate just massless spin-two particles. An extension of the analysis of \cite{Boulanger:2000rq} to the
transverse subgroup is currently under investigation \cite{BD} {\bf (see also \cite{Blas:2007pp})}.}.
\\

Let us now consider the case of generic $m_s$ and $m_t$.
For simplicity we will concentrate on perturbations around de
Sitter solutions which will be foliated by spatially flat
sections,
\be
\label{deSitter}
\Omega_{\mu\nu}\d x^\mu \d x^\nu=a(\eta)^2(\d \eta^2
-\delta_{ij} \d x^i\d x^2),
\ee
where $a(\eta)=-(H\eta)^{-1}$, $H^2=\Lambda_g/3$ being a constant
and $\eta\in (-\infty, 0)$. The kinetic term in (\ref{action})
will be given by (cfr. (\ref{lgenerbi}))
\ba
L_K\equiv-\frac{1}{2\kappa_g} \sqrt{-g}\ (R_g+2\Lambda_g)
-\frac{1}{2\kappa_f} \sqrt{-f}\ (R_f+2\Lambda_f),
\ea
with $\L_f=\g^{-1}\L_g$. To second order in perturbations we can rewrite the kinetic term
in terms of a massive and a massless field,
\ba
\label{massivemasslesss}
L_K=-\frac{1}{2\kappa_+} \sqrt{-g_+}\ (R_{g_+}+2\Lambda_g)
-\frac{1}{2\kappa_-} \sqrt{-g_-}\ (R_{g_-}+2\Lambda_g)+o(h^3),
\ea
where $\kappa_-=\frac{\kappa_g}{1+\kappa}$,
 $\kappa_+=\kappa_g\kappa^{-1}(1+\kappa)$, with $\kappa=\g \kappa_g \kappa_f^{-1}$,
  $g_-{}_{\mu\nu}=\Omega_{\mu\nu}+h^-_{\mu\nu}$ and $g_+{}_{\mu\nu}=\Omega_{\mu\nu}
+ h^+_{\mu\nu}$ . Besides, we have introduced
the  massive and  massless combinations
\be
\label{massless}
 h^+_{\m\n}=h^g_{\m\n}+h^f_{\m\n},\quad h^-_{\m\n}=(1+\kappa)^{-1}\left(h^g_{\m\n}-\kappa h^f_{\m\n}\right).
\ee
The dynamics of the massless part is well known. One easily finds
that only the tensor modes are dynamical.
For the generic massive theory in de Sitter space,
studying the longitudinal mode
of the massive representation we would argue that the only ghost-free possibility
is the Fierz-Pauli mass term, $m^2_t=-m^2_s$ \cite{Fierz:1939ix,Arkani-Hamed:2002sp}.
However, in general, this mode decouples only at high energies (larger than a combination of
the rest of relevant mass scales). For intermediate energy scales, the longitudinal mode is
coupled to another scalar mode which can modify this picture \cite{Dubovsky:2004sg,
Creminelli:2005qk}. Also, the curvature scale $H$ could play a
role in making these intermediate scales phenomenologically relevant.
We will study this possibility directly in the unitary gauge.

Let us
first split the degrees of freedom of the massive combination into scalar,
vectorial and tensorial modes,
\ba
h^+_{00}&=&2a(\eta)^2 A,\nonumber\\
h^+_{0i}&=&a(\eta)^2(B_{,i}+V_i),\nonumber \\
h^+_{ij}&=&a(\eta)^2(2\psi \delta_{ij}-2 E_{,ij}-2 F_{(i,j)}-t_{ij}),
\label{decomp}
\ea
where $\psi$, $B$, and $E$ are the scalar
modes, $F_i$ and $V_i$ are vector modes, and $t_{ij}$ is a tensor
mode. The vector modes are divergenceless and the tensor modes are
transverse and traceless.

The expansion of the kinetic term in this foliation can be extracted
from the usual expansion in de Sitter space (see e.g. \cite{Mukhanov:1990me},
notice however the difference of convention). One finds
\ba
\label{kinetic}
-\frac{1}{2\kappa_+} \int \d^4 &x&\sqrt{-g_+} (R_++2\Lambda_g)=
-\frac{1}{2\kappa_+}\int \d^4 x a^2(\eta)\Big\{\frac{1}{4}t_{ij}\Box t_{ij} +\frac{1}{2}(V_i+F'_i)\Delta (V_i+F'_i)\nonumber\\
&&+6(\psi'+\H A)^2-2 \Delta \psi(2A-\psi)-4\Delta(B+E')(\psi'+\H A)
\Big\},
\ea
where $\H=a(\eta)'/a(\eta)=a(\eta)H$ and the prime refers to derivative with respect to the conformal time $\eta$.
We have also introduced the d'Alembertian $\Box=\eta^{\m\n}\pd_\m \pd_\n$ and
the Laplacian $\triangle=\pd_i \pd_i$.
 The interaction term (\ref{lgenerbi}) reads
\ba
\label{linter}
\tilde L_{int}&=& \frac{1}{2\kappa_+}a(\eta)^4\Big\{m^2_s(A+\Delta E-3\psi)^2\nonumber\\
&&-\frac{1}{4}m^2_t\Big(t_{ij}t_{ij}-2(V_i V_i+F_i \Delta F_i)
 +4 (A^2+\frac{B \Delta B}{2}+(\Delta E)^2+3\psi^2-2\psi \Delta E)\Big)\Big\}.
\ea
We can now analyse the different components in turn.

 \subsection{Tensor and Vector Modes}

The action for the massive tensorial modes is simply
\be
{}^{(t)}\delta S_2=-\frac{1}{8\kappa_+}\int \d x^4  a^2(\eta)\Big(t_{ij}\Box
t_{ij}+a(\eta)^2m^2_t  t_{ij}t_{ij}\Big).
\ee
From
this equation we can read the mass of the graviton which will be
given by
$ m^2_t,$
and the tachyon-free condition will simply read
$$m^2_t\geq 0.$$
Regarding the vector modes, their action is
\be
{}^{(v)}\delta S_2=-\frac{1}{4\kappa_+}\int \d x^4  a^2(\eta)\Big((V_i+ F'_i)\Delta
(V_i+ F'_i)-a^2(\eta)m^2_t (V_i V_i+F_i\Delta F_i)\Big).
\ee
The field
$V_m$ enters the action
without time derivatives, and thus its variation yields the constraint,
\be
\triangle (V_i+ F'_i)= a(\eta)^2 m^2_t
V_i\equiv m^2(\eta) V_i.
\ee
Taking this constraint into account, the action for the
vectorial modes up to second order can be written as
\be
{}^{(v)}\delta
S_2= \frac{1}{4\kappa_+} \int \d^4 x a^2(\eta)m^2(\eta)\Big( F'_i
\frac{\Delta}{\Delta-m(\eta)^2}  F'_i +F_i\Delta F_i\Big)
\ee
This Lagrangian has the usual signs, and thus no ghost or tachyons appear in
the theory for $m^2_t\geq 0$. More concretely, we can canonically
normalize the previous field equation with the field redefinition
\be
F_i^c=m(\eta)\sqrt{\frac{\Delta}{\kappa(\Delta-m(\eta)^2)}}F_i.
\ee
We conclude that the only constraint
we get form the analysis of the vector and tensor modes is $m^2_t\geq0$.

\subsection{Scalar Modes}
From (\ref{kinetic}) and (\ref{linter}), the second
order Lagrangian for the scalar part reads
\ba
{}^{(2)}\delta S_2&&= \frac{1}{2\kappa_+}\Big[\int \d^4 x a^2(\eta)\{
-6( \psi'+ {\mathcal{H}}A)^2
+2 \Delta \psi(2A-\psi)+4\Delta(B+E')( \psi'+{\mathcal{H}}A)\}\nonumber\\
&&+\int\d^4x a^4(\eta)\Big(m^2_s(A+\Delta E-3\psi)^2-m^2_t\{3 \psi^2
+(\Delta E)^2 -2 \psi \Delta E+ \frac{B\Delta B}{2}+A^2\}\Big)\Big].
\ea
$B$ is non-dynamical, and for $m^2_t\neq 0$ it is determined in terms of
the other fields.
For $m^2_t=m^2_s$, $A$ appears only linearly in the mass term. For the flat case $H=0$ and $a(\eta)=1$,
this makes $A$ a Lagrange multiplier
and thus its variation gives rise to a constraint between the fields
$E$ and $\psi$, leaving just one scalar propagating degree of freedom. In the
de Sitter case, the result is the same, although this is not so obvious from the previous expression
for the action until one substitutes the constraints. \\

The variation with respect to $A$ and $B$  yields the
constraints
\ba
B&=&\frac{4(\psi'+{\mathcal{H}}A)}{ a(\eta)^2 m^2_t},\\
A&=&\frac{ -2a(\eta)^2 m^2_t ({\mathcal{H}}(\phi'-3\psi')
+\Delta\psi)- a(\eta)^4 m^2_s  m^2_t (\phi-3 \psi) -8\Delta{\mathcal{H}}
\psi'}{  m^2_t( m^2_s- m^2_t)a(\eta)^4+8\Delta \H^2-6  m^2_t a(\eta)^2 \H^2},
\ea
where $\phi=\Delta E$. Let us first
consider the kinetic part of the action, which after insertion of the constraints
reads
\ba
K=
\frac{a(\eta)^2}{2\kappa_+}(M_1(\eta) \phi \psi'
+M_2(\eta) \psi'^2+M_3(\eta) \psi'\phi'+M_4(\eta) \phi'^2),
\ea
where we have performed a partial integration to eliminate the term $\phi'\psi$.
The
functions $M_i(\eta)$ are given by
\ba
M_1(\eta)&=&\frac{ 8\Delta (m^2_t-2  m^2_s) a(\eta)^2 \H}
{ m^2_t( m^2_s- m^2_t)a(\eta)^4+8\Delta \H^2-6  m^2_t a(\eta)^2 \H^2},\\
M_2(\eta)&=&\frac{2 ( m^2_s-m^2_t)a(\eta)^2(4\Delta -3m^2_t a(\eta)^2)}
{ m^2_t( m^2_s- m^2_t)a(\eta)^4+8\Delta \H^2-6  m^2_t a(\eta)^2 \H^2},\\
M_3(\eta)&=&\frac{4 m^2_t(m^2_s-m^2_t) a(\eta)^4}
{ m^2_t( m^2_s- m^2_t)a(\eta)^4+8\Delta \H^2-6  m^2_t a(\eta)^2 \H^2},\\
M_4(\eta)&=&\frac{-4 m^2_t a(\eta)^2 \H^2}{ m^2_t( m^2_s- m^2_t)a(\eta)^4
+8\Delta \H^2-6  m^2_t a(\eta)^2 \H^2}.
\ea
A difference
between the flat and the de Sitter backgrounds is that the coefficients
$M_1(\eta)$ and $M_4(\eta)$ cancel in the former case, and this automatically yields a kinetic
term with a negative
eigenvalue unless $M_3(\eta)=0$ which happens for the Fierz-Pauli combination $m_s^2=m_t^2$.
The situation in de Sitter is slightly more complicated.
\\

Let us now show that the previous kinetic term gives a positive contribution to the
Hamiltonian in the range of parameters
\ba
&&m^2_t \geq 0, \quad 0\leq m^2_s -m^2_t \leq 6 H^2 \label{noghostI}.
\ea
Indeed, the kinetic term can written as
\be
K=\frac{a(\eta)^2}{2\kappa_+}\left(M_1(\eta)\phi\psi'+\left(M_2(\eta)
-\frac{M_3^2(\eta)}{4M_4(\eta)}\right)\psi'^2+M_4(\eta)\left(
\phi'+\frac{M_3(\eta)}{2M_4(\eta)}\psi'\right)^2\right).
\ee
In the range (\ref{noghostI}), $M_4(\eta)$ and
$4 M_4(\eta)M_2(\eta)-M_3^2(\eta)$ are positive. By Euler's theorem,
the corresponding Hamiltonian
\ba
\label{hamiltonian}
{\mathrm H}_K\equiv
\Pi_\phi  \phi'+\Pi_\psi  \psi'-K,
\ea
is numerically equal to the two last terms in the Lagrangian, which are quadratic in
generalized velocities, and hence it is positive definite.
The second condition in (\ref{noghostI}) for a positive kinetic term
reduces to the usual $m^2_s=m^2_t$ for the
Minkowski limit $H=0$. For  $H>0$ the endpoints of the interval
are of different nature: the condition
$m^2_s-m^2_t\geq 0$ is necessary condition for positivity of $M_2-M_3^2/M_4$ at any value of
the momentum, whereas the upper bound on the range of $m_s^2-m_t^2$ can be somewhat relaxed
depending on the value of the momentum. Indeed, what we need is
that
\be
m_s^2-m^2_t \leq 6 H^2 \left[ 1-{4\Delta\over 3a^2m^2_t}\right],
\ee
so the condition is considerably relaxed at wavelengths shorter than the inverse graviton mass.\\

Once we have established the positivity of part of the Hamiltonian, let us see what happens
to rest of it, namely to the potential part. This part will be given by
\be
V\equiv K-L
=\frac{a(\eta)^2}{2\kappa_+}(M_5(\eta)\phi^2
+M_6(\eta)\phi\psi+M_7(\eta)\psi^2),
\ee
where the coefficients are rather cumbersome and we omit them.
Before proceeding, it should be noted that the Hamiltonian we are
considering is time dependent, and hence not conserved. Its positivity
and boundedness is a useful criterion only as long as we consider time-scales
shorter than the expansion time, or energies larger than $H$. This is what
we may call the adiabatic limit. Hence, let us assume that $m_s,m_t \gg H$,
even if their difference is much smaller $m_s^2 - m_t^2 \lesssim H^2$, so that we
can satisfy the positivity of the kinetic term as discussed above.
We have checked that within this adiabatic limit, the potential $V$ grows negative
and unbounded below for $-\Delta/a^2 \gg m^2$.
Instabilities at high momenta have been previously studied in \cite{Dubovsky:2005xd}, and they are just
as bad as ghost instabilities. Unlike the case of tachyons, the phase space for
instability is infinite and this yields infinite decay rates.

If the masses $m_t$ and $m_s$ are small, of order of the expansion rate $H$, then
we are outside of the adiabatic limit, and the Hamiltonian above is
not a very useful indicator of stability. Instead, we should use a conserved charge
associated to the time-like Killing vector for length scales smaller than
the horizon \cite{Abbott:1981ff}. Due to the existence of the cosmological scale,
it is in principle possible (although by no means clear) that there may be some range
\be
\label{range}
m^2 \left(\frac{H}{m}\right)^{\a}\gtrsim -\Delta/a^2\gtrsim H^2 \gtrsim m^2,
\ee
(with $\alpha>2$), where this conserved charge is positive definite.
The effective theory would then be well defined for momenta larger than $H$
(corresponding to modes within the horizon), provided that the theory is cut-off at the energy
scale $m(H/m)^{\alpha/2}$. We leave the study of this conserved charge for further research.
We note, however, that we need a theory which is applicable to wavelenghs {\em much} smaller than
the horizon $-\Delta/a^2\gg H^2$, where the adiabatic approximation should again be valid.
We have checked that for $-\Delta/a^2\gg H^2 \gtrsim m^2$, the potential $V$ grows negative and
unbounded below, so the possibility of a range of the form
(\ref{range}) where the conserved charge is positive does not look particularly promising.
\\

Finally, for the case $m^2_s=m^2_t$ the analysis of the degrees of freedom
has already been performed in another foliation in \cite{Deser:2001wx} (see
also \cite{Bengtsson:1994vn}). In our analysis for this case we find
 $M_2(\eta)=M_3(\eta)=0$ and thus $\psi$ is not a
propagating field. After varying the action with respect to
$\psi$ we obtain a constraint which after substitution yields the Lagrangian
\ba
\Upsilon\Big( \phi'^2
+\frac{9m^4_t a(\eta)^4\m_H-21 m^2_t a(\eta)^2\mu_H\Delta+4(\m_H-6m^2_t a(\eta)^2)\Delta^2+2\Delta^3}{(9m^2_t a(\eta)^2\m_H-
6\m_H\Delta -2\Delta^2)}\ \phi^2\Big),\nonumber
\ea
where $2\mu_H=2\H^2-m^2_t a(\eta)^2$ and
$$
\Upsilon=\frac{3a(\eta)^4m^2_t \mu_H}{\kappa_+ (9m^2_t a(\eta)^2\m_H-
6\m_H\Delta -2\Delta^2)}.
$$
This Lagrangian will be ghost-free and tachyon-free for $\m_H\leq0$. This reduces to
the well known condition $m^2\geq 2H^2$ \cite{Higuchi:1986py}.

Another case which differs from the usual Fierz-Pauli
term and is still well defined is the case of Lorentz-breaking mass terms
\cite{Rubakov:2004eb,Dubovsky:2004sg}. These mass terms arise
from  solutions of the form sketched in Section 2. A simple example would be given by
two de Sitter metrics with $\l_i=\{\b\g^{-1},\g^{-1},\g^{-1},\g^{-1}\}$.
The interesting ranges propagating only one scalar or no scalars at all are the same as those in
\cite{Rubakov:2004eb,Dubovsky:2004sg}, and for generic mass terms with two scalars, there is
always a gradient instability at high energies as happens in the Lorentz preserving case \cite{BD}.

\section{Conclusions}

Bigravity is a natural arena where a non-linear theory of massive gravity can be
formulated. This theory is invariant under the ``diagonal" group of diffeomorphisms, under
which both metrics transform.

We have shown that if the interaction between the two metrics is non-derivative, we can always find exact
solutions in the Schwarzschild-(A)dS family, with Lorentz breaking asymptotics.
These are ``Type I" solutions \cite{Isham:1977rj},
in which both metrics cannot (generically) be brought to diagonal form in the same coordinate system.
For a given interaction potential, the degree of Lorentz-breaking is an adjustable parameter.

Perturbations around Lorentz-breaking bi-flat (or bi-de Sitter \cite{BD})
solutions lead to gravitons with Lorentz-breaking mass terms. Because of the invariance under
diagonal diffeomorphisms \cite{Berezhiani:2007zf}, mass terms with components $h_{0i}$ are absent from the
second order Lagrangian. This, in turn, leads to a well behaved theory of linearized perturbations
\cite{Berezhiani:2007zf,Gabadadze:2004iv},
which is not afflicted by the vDVZ discontinuity. It is somewhat puzzling that in the linear theory,
there are corrections to the Newtonian potential which are proportional to the square of the graviton mass
and which grow linearly with the distance to the origin. On the other hand, as mentioned above,
these theories admit the Schwarzschild metric as an exact solution for the same values of the parameters.
Thus, the linearized solutions for a static spherically symmetric sources do not coincide with the
linearization of the known vacuum solutions. This seems to indicate that this theory has a linearization
instability such as the one which is found in other contexts \cite{Moncrief:1976un}, some of which
are related to massive gravity and may have important phenomenological
consequences \cite{Deffayet:2006wp}. Another possibility is that
there may be other exact solutions which coincide with the linearized approximation at large distances, and
those may be the relevant ones which can be matched to spherically symmetric matter sources near the origin.
This issue clearly deserves further investigation.

We have also considered perturbations to solutions where both metrics are proportional to each other, focusing
in the case of de Sitter. This has led us to consider generic mass terms for gravitons in de Sitter space, beyond
the Fierz-Pauli case. We find that for the case of Lorentz-invariant mass terms, only the Fierz-Pauli combination
is free from instabilities at high momenta $-\Delta/a^2 \gg  m^2$  (and only for $m^2\geq2H^2$). For the
Lorentz-breaking case, the situation is analogous to that in flat space \cite{Rubakov:2004eb,Dubovsky:2004sg,BD}.

\section*{Acknowledgements}
It is a pleasure to thank Alberto Iglesias, Oriol Pujolas and Joan Soto for useful discussions.
D.B. thanks the Institut d'Astrophysique de Paris for its hospitality, while this work
was being concluded. The work of D.B. has been supported by MEC (Spain) through a FPU grant.
J.G. and D.B. acknowledge support from CICYT grant FPA 2004-04582-C02-02and DURSI 2005SGR 00082.



\begin{thebibliography}{99}

\bibitem{Albrecht:2006um}
  A.~Albrecht {\it et al.},
  arXiv:astro-ph/0609591.



\bibitem{Fierz:1939ix}
  M.~Fierz and W.~Pauli,
  Proc.\ Roy.\ Soc.\ Lond.\  A {\bf 173} (1939) 211.


\bibitem{vDVZ}
H.~van Dam and M.~J.~Veltman,
Nucl.\ Phys.\ B {\bf 22} (1970) 397.
V.I.Zakharov, JETP Lett {\bf 12}, 312 (1970).
Y.~Iwasaki,
Phys.\ Rev.\ D {\bf 2} (1970) 2255.

\bibitem{Dvali:2000hr}
  G.~R.~Dvali, G.~Gabadadze and M.~Porrati,
  Phys.\ Lett.\  B {\bf 485} (2000) 208
  [arXiv:hep-th/0005016].

\bibitem{Deffayet:2000uy}
  C.~Deffayet,
  Phys.\ Lett.\  B {\bf 502} (2001) 199
  [arXiv:hep-th/0010186].
  C.~Deffayet, G.~R.~Dvali and G.~Gabadadze,
  Phys.\ Rev.\  D {\bf 65} (2002) 044023
  [arXiv:astro-ph/0105068].

\bibitem{Damour:2002ws}
  T.~Damour and I.~I.~Kogan,
  Phys.\ Rev.\ D {\bf 66} (2002) 104024
  [arXiv:hep-th/0206042].


\bibitem{Damour:2002wu}
  T.~Damour, I.~I.~Kogan and A.~Papazoglou,
  Phys.\ Rev.\  D {\bf 66} (2002) 104025
  [arXiv:hep-th/0206044].


\bibitem{Weinberg:1988cp}
  S.~Weinberg,
  Rev.\ Mod.\ Phys.\  {\bf 61} (1989) 1.

\bibitem{Vainshtein:1972sx}
  A.~I.~Vainshtein,
  Phys.\ Lett.\  B {\bf 39} (1972) 393.

\bibitem{Jun:1986hg}
  J.~H.~Jun and I.~Kang,
  Phys.\ Rev.\  D {\bf 34} (1986) 1005.

\bibitem{Damour:2002gp}
  T.~Damour, I.~I.~Kogan and A.~Papazoglou,
  Phys.\ Rev.\  D {\bf 67} (2003) 064009
  [arXiv:hep-th/0212155].





\bibitem{Deffayet:2001uk}
  C.~Deffayet, G.~R.~Dvali, G.~Gabadadze and A.~I.~Vainshtein,
  Phys.\ Rev.\ D {\bf 65} (2002) 044026
  [arXiv:hep-th/0106001].


\bibitem{Gruzinov:2001hp}
  A.~Gruzinov,
  New Astron.\  {\bf 10} (2005) 311
  [arXiv:astro-ph/0112246].
  A.~Lue,
  Phys.\ Rev.\  D {\bf 66}, 043509 (2002)
  [arXiv:hep-th/0111168].
  M.~Porrati,
  Phys.\ Lett.\  B {\bf 534}, 209 (2002)
  [arXiv:hep-th/0203014].
  T.~Tanaka,
  Phys.\ Rev.\  D {\bf 69} (2004) 024001
  [arXiv:gr-qc/0305031].
  G.~Gabadadze and A.~Iglesias,
  Phys.\ Rev.\  D {\bf 72}, 084024 (2005)
  [arXiv:hep-th/0407049].

\bibitem{Dvali:2006if}
  G.~Dvali, G.~Gabadadze, O.~Pujolas and R.~Rahman,
  arXiv:hep-th/0612016.



\bibitem{Rubakov:2004eb}
  V.~A.~Rubakov,
  arXiv:hep-th/0407104.



\bibitem{Dubovsky:2004sg}
  S.~L.~Dubovsky,
  JHEP {\bf 0410} (2004) 076
  [arXiv:hep-th/0409124].

\bibitem{Libanov:2005nv}
  M.~V.~Libanov and V.~A.~Rubakov,
  Phys.\ Rev.\  D {\bf 72} (2005) 123503
  [arXiv:hep-ph/0509148].

\bibitem{Gripaios:2004ms}
  B.~M.~Gripaios,
  JHEP {\bf 0410} (2004) 069
  [arXiv:hep-th/0408127].

\bibitem{Dvali:2007ks}
  G.~Dvali, O.~Pujolas and M.~Redi,
  arXiv:hep-th/0702117.

\bibitem{ArkaniHamed:2003uy}
  N.~Arkani-Hamed, H.~C.~Cheng, M.~A.~Luty and S.~Mukohyama,
  JHEP {\bf 0405} (2004) 074
  [arXiv:hep-th/0312099].


\bibitem{Blas:2005yk}
  D.~Blas, C.~Deffayet and J.~Garriga,
  Class.\ Quant.\ Grav.\  {\bf 23} (2006) 1697
  [arXiv:hep-th/0508163].

\bibitem{Berezhiani:2007zf}
  Z.~Berezhiani, D.~Comelli, F.~Nesti and L.~Pilo,
  arXiv:hep-th/0703264.


\bibitem{Blas:2005P}
  D.~Blas,
  Proceedings of Peyresq Cosmology Meeting 2005,
      Int.\ J.\ Theor.\ Phys.\
      \emph{In press} (for an on-line version see
      http://www.springerlink.com/content/k46748731n22v526/).


\bibitem{Blas:2005sz}
  D.~Blas,
  AIP Conf.\ Proc.\  {\bf 841} (2006) 397.



\bibitem{Isham:gm}
C.~J.~Isham, A.~Salam and J.~Strathdee,
Phys.\ Rev.\ D {\bf 3} (1971) 867.

\bibitem{Boulware:1973my}
  D.~G.~Boulware and S.~Deser,
  Phys.\ Rev.\  D {\bf 6}, 3368 (1972).

\bibitem{Creminelli:2005qk}
  P.~Creminelli, A.~Nicolis, M.~Papucci and E.~Trincherini,
  JHEP {\bf 0509} (2005) 003
  [arXiv:hep-th/0505147].

\bibitem{Isham:1977rj}
C.~J.~Isham and D.~Storey,
Phys.\ Rev.\ D {\bf 18}, 1047 (1978).



\bibitem{Deffayet:2003zm}
  C.~Deffayet and J.~Mourad,
Class.\ Quant.\ Grav.\  {\bf 21} (2004) 1833
  [arXiv:hep-th/0311125].
  C.~Deffayet and J.~Mourad,
  Phys.\ Lett.\  B {\bf 589} (2004) 48
  [arXiv:hep-th/0311124].


\bibitem{Dubovsky:2004ud}
  S.~L.~Dubovsky, P.~G.~Tinyakov and I.~I.~Tkachev,
  Phys.\ Rev.\ Lett.\  {\bf 94} (2005) 181102
  [arXiv:hep-th/0411158].


\bibitem{Gabadadze:2004iv}
  G.~Gabadadze and L.~Grisa,
  Phys.\ Lett.\  B {\bf 617} (2005) 124
  [arXiv:hep-th/0412332].

\bibitem{Dubovsky:2005dw}
  S.~L.~Dubovsky, P.~G.~Tinyakov and I.~I.~Tkachev,
  Phys.\ Rev.\ D {\bf 72} (2005) 084011
  [arXiv:hep-th/0504067].

\bibitem{Babichev:2006vx}
  E.~Babichev, V.~F.~Mukhanov and A.~Vikman,
  JHEP {\bf 0609} (2006) 061
  [arXiv:hep-th/0604075].


\bibitem{Arkani-Hamed:2003uy}
N.~Arkani-Hamed, H.~C.~Cheng, M.~A.~Luty and S.~Mukohyama,
JHEP {\bf 0405} (2004) 074
[arXiv:hep-th/0312099].



\bibitem{Deffayet:2005ys}
  C.~Deffayet and J.~W.~Rombouts,
  Phys.\ Rev.\  D {\bf 72} (2005) 044003
  [arXiv:gr-qc/0505134].



\bibitem{Stelle:1977ry}
  K.~S.~Stelle,
  Gen.\ Rel.\ Grav.\  {\bf 9} (1978) 353.

\bibitem{Alvarez:2006uu}
  E.~Alvarez, D.~Blas, J.~Garriga and E.~Verdaguer,
  Nucl.\ Phys.\  B {\bf 756} (2006) 148
  [arXiv:hep-th/0606019].

\bibitem{Boulanger:2000rq}
  N.~Boulanger, T.~Damour, L.~Gualtieri and M.~Henneaux,
  Nucl.\ Phys.\ B {\bf 597} (2001) 127
  [arXiv:hep-th/0007220].



\bibitem{BD}
D.~Blas \emph{In progress.}

\bibitem{Blas:2007pp}
  D.~Blas,
  J.\ Phys.\ A  {\bf 40} (2007) 6965
  [arXiv:hep-th/0701049].

\bibitem{Arkani-Hamed:2002sp}
N.~Arkani-Hamed, H.~Georgi and M.~D.~Schwartz,
Annals Phys.\  {\bf 305} (2003) 96 [arXiv:hep-th/0210184].




\bibitem{Mukhanov:1990me}
  V.~F.~Mukhanov, H.~A.~Feldman and R.~H.~Brandenberger,
  Phys.\ Rept.\  {\bf 215} (1992) 203.

\bibitem{Dubovsky:2005xd}
  S.~Dubovsky, T.~Gregoire, A.~Nicolis and R.~Rattazzi,
  JHEP {\bf 0603} (2006) 025
  [arXiv:hep-th/0512260].

\bibitem{Abbott:1981ff}
  L.~F.~Abbott and S.~Deser,
  Nucl.\ Phys.\  B {\bf 195} (1982) 76.

\bibitem{Deser:2001wx}
  S.~Deser and A.~Waldron,
  Phys.\ Lett.\ B {\bf 508} (2001) 347
  [arXiv:hep-th/0103255].

\bibitem{Bengtsson:1994vn}
  I.~Bengtsson,
  J.\ Math.\ Phys.\  {\bf 36} (1995) 5805
  [arXiv:gr-qc/9411057].

\bibitem{Higuchi:1986py}
  A.~Higuchi,
  Nucl.\ Phys.\ B {\bf 282} (1987) 397.



\bibitem{Moncrief:1976un}
  V.~Moncrief,
  J.\ Math.\ Phys.\  {\bf 17} (1976) 1893.
  D.~Kastor and J.~H.~Traschen,
  Phys.\ Rev.\  D {\bf 47} (1993) 480.
  A.~Higuchi,
Class.\ Quant.\ Grav.\  {\bf 8} (1991) 2023.

\bibitem{Deffayet:2006wp}
  C.~Deffayet, G.~Gabadadze and A.~Iglesias,
  JCAP {\bf 0608} (2006) 012
  [arXiv:hep-th/0607099].





\end{thebibliography}
\end{document}